\documentclass{aa}
\usepackage{natbib} 
\usepackage[pdftex]{graphicx} 
\usepackage{booktabs} 

\bibpunct{(}{)}{;}{a}{}{,} 
\usepackage{amsmath}
\usepackage{caption}
\usepackage{braket}
\usepackage{bm}
\usepackage[varg]{txfonts}

\usepackage{color}

\begin{document} 

\title{An updated maximum likelihood approach to open cluster distance determination}
\author{M. Palmer \inst{1}
\and F. Arenou \inst{2}
\and X. Luri \inst{1}
\and E. Masana\inst{1}}
\institute{Dept. d'Astronomia i Meteorologia, Institut de Ci\`{e}ncies del Cosmos, Universitat de Barcelona (IEEC-UB), Mart\'{i} Franqu\`{e}s 1, E08028 Barcelona, Spain.\label{1}
\and GEPI, Observatoire de Paris, CNRS, Universit\'{e} Paris Diderot, 5 place Jules Janssen, 92190, Meudon, France \label{2}}


\abstract{}
{An improved method for estimating distances to open clusters is presented and applied to Hipparcos data for the Pleiades and the Hyades. The method is applied in the context of the historic Pleiades distance problem, with a discussion of previous criticisms of Hipparcos parallaxes. This is followed by an outlook for Gaia, where the improved method could be especially useful.}
{Based on maximum likelihood estimation, the method combines parallax, position, apparent magnitude, colour, proper motion, and radial velocity information to estimate the parameters describing an open cluster precisely and without bias. }
{We find the distance to the Pleiades to be $120.3 \pm 1.5$ pc, in accordance with previously published work using the same dataset. We find that error correlations cannot be responsible for the still present discrepancy between Hipparcos and photometric methods. Additionally, the three-dimensional space velocity and physical structure of Pleiades is parametrised, where we find strong evidence of mass segregation. The distance to the Hyades is found to be $46.35\pm 0.35$ pc, also in accordance with previous results. Through the use of simulations, we confirm that the method is unbiased, so will be useful for accurate open cluster parameter estimation with Gaia at distances up to several thousand parsec.}
{}
\maketitle

\section{Introduction}

Open clusters have long been used as a testing ground for a large number of astronomical theories. Determining the distances to nearby open clusters is critical, since they have historically formed the first step in the calibration of the distance scale. Because all stars within an open cluster are expected to have the same age and metallicity, accurate distance estimates are highly useful in calibrating the main sequence and checking stellar evolutionary theory through comparison with theoretical isochrones. 

Until recently, accurate distances to even the most nearby clusters have not been possible. The Hipparcos astrometric mission of 1989 \citep{Hipparcos} for the first time gave accurate parallax measurements for over one hundred thousand stars and has been used extensively to give direct distance measurements to more than 30 open clusters.

Still, many questions remain. While revolutionary in its time, the milliarcsecond astrometry and limiting magnitude ($H_p$\textless$12.5$) of Hipparcos allow calculating distances to only the nearest open clusters, and even then do so with a precision no better than a few percentage points. Owing to the relatively small size of most open clusters compared with their distances and the precision of measurements, the Hipparcos data has been unable to give definitive answers about the internal structure and physical size of such clusters. 

Additionally, the release of the Hipparcos catalogue in 1997 has led to some controversy. The most famous is the case of the Pleiades, where methods based on Hipparcos data (\cite{FVL1997}, \cite{robichon}, \cite{mermilliod}) gave a distance estimate that was some 10\% shorter than works based on photometric methods (\cite{MSfitting1}, \cite{MSfitting2}, \cite{photometricDistance}) (see Sect. \ref{sec:correlations}). 

With the launch of Gaia, the European Space Agency's second major astrometric satellite, the situation is set to change. Building on the success of Hipparcos, Gaia will provide micro-arcsecond astrometric precision and a limiting magnitude of 20, which will revolutionise many aspects of astronomy. 

With the above in mind, it is apparent that a new method is required that is capable of squeezing the maximum precision from the currently available data and is capable of utilising information from the full range of observed quantities (astrometric, photometric, and kinematic information). This will be particularly true after the launch of Gaia, which will produce a rich dataset that will include not only accurate parallax measurements, but also photometry at millimag precision and a full set of kinematics obtained from proper motions combined with radial velocity measurements from the on-board radial velocity spectrometer (for stars with $G_{RVS}$\textless$17$). 

The use of trigonometric parallaxes can be problematic, and care must be taken to account for effects such as the \cite{lutzkelker} or \cite{malmquist} biases, sample selection effects and non-linear transformations, such as those highlighted by \cite{Arenou2} and \cite{Arenou1}.

In Sect. \ref{sec:methods} the rationale behind the method is described, followed by the exact mathematical formulation in Sect. \ref{sec:math}. A description of the observational data used is given in Sect. \ref{sec:data}, and the results of application of the method to the Pleiades and Hyades given in Sects. \ref{sec:pleiades} and \ref{sec:hyades}. The possible effects of correlated errors in the Hipparcos catalogue are discussed in Sect. \ref{sec:correlations}. The use of the method with Gaia data is tested using simulations in Sect. \ref{sec:OutlookForGaia}. 

\section{Methods}
\label{sec:methods}

Maximum likelihood estimation (MLE) has been used in relation to open clusters since 1958, where \cite{v1958} used MLE to perform cluster membership selection from proper motions. 

For the Hyades and Pleiades, Chen and Zhao used a combination of the convergent point method with MLE in order to simultaneously determine mean parallax and kinematics for the Hyades \citep{CZHyades} and later the Pleiades \citep{CZPleiades}. These early approaches applied the principles of MLE in its basic form, but improvements by \cite{luri1999} resolved several shortcomings in its use. The improvements, described in full in \cite{luri1999}, have been used extensively in this work and allow the following improvements:
\begin{itemize}
\item{The use of numerical methods avoids the necessity of approximating or simplifying complex equations during the formulation of the MLE, thus avoiding any loss of precision.}
\item{Explicitly taking into account sample selection effects caused by observational constraints is required to correctly model the joint probability density function (PDF) of the sample. These selection effects introduce various biases into the sample, e.g. Malmquist, and their correct treatment within the formulation of the density laws avoids the need for a posteriori corrections, which are often poorly understood.}
\item{Through the construction of the MLE joint PDF, where all of a star's available information is used simultaneously, posterior estimates of a star's properties can be obtained with increased precision compared with the original data. This can be extended to calculate cluster membership probabilities without the need for external membership selection.}
\item{The three-dimensional spatial distribution of the stars in a cluster is modelled, along with their kinematic distribution, assuming all the members follow a single velocity ellipsoid. Additionally, their absolute magnitude is modelled assuming the stars absolute magnitudes can be described by an isochrone. This links each of a star's observables to the properties of the cluster to which it belongs, giving tight constraints on the quantification of a clusters parameters. With a sufficient quantity of data, it is possible to extract higher order information, as can be achieved through a Bayesian hierarchical model.}
\end{itemize}

\section{Mathematical formulation}
\label{sec:math}
\subsection{Definition}
The MLE is a method for estimating the parameters of a statistical model. By specifying a joint PDF for all observables and taking an initial guess at the value of each parameter, the likelihood of having taken a set of observations is calculated. Maximising this function by varying the parameters results in the MLE of the parameters in the statistical model.

The likelihood function can be defined as

\begin{equation}
\mathcal{V}(\boldsymbol{\theta}) = \prod\limits_{i=1}^n \mathcal{O}(\mathbf{y}_i|\boldsymbol{\theta}) 
\label{eqn:MLBasic}
\end{equation}
where the joint PDF $\mathcal{O}(\mathbf{y}_i|\boldsymbol{\theta})$ is made up of the un-normalised PDF $\mathcal{D}(\mathbf{y}_i|\boldsymbol{\theta})$ and a normalisation constant $\mathcal{C}_i$, such that
\begin{equation} 
\mathcal{O}(\mathbf{y}_i|\boldsymbol{\theta}) = \mathcal{C}_i^{-1} \mathcal{D}(\mathbf{y}_i|\boldsymbol{\theta})
,\end{equation} 
and $\boldsymbol{\theta}$ is the vector giving the parameters of the model. 

In MLE, the user is free to define any model. For reliable results, the models chosen must resemble the system being modelled, and can incorporate the a priori information one may have about the system.
In our case, we have chosen that $\boldsymbol{\theta} = (R,\sigma_R,M(V-I),\sigma_M,U,V,W,\sigma_{UVW})$, where $R$ is the distance to the centre of the cluster assuming a spherical Gaussian distribution, $\sigma_R$  the intrinsic dispersion around the mean distance, $M(V-I)$  the mean absolute magnitude as a function of colour, $\sigma_M$  the intrinsic dispersion around the mean absolute magnitude, $U$, $V$, and $W$ are the three components of the clusters velocity ellipsoid in galactic Cartesian coordinates, and $\sigma_{UVW}$ is the intrinsic dispersion in the clusters velocity. 

The vector $(\mathbf{y}=m,l,b,\varpi,\mu_l,\mu_{\delta},v_r)$ describes the observed properties of each star, and $(\mathbf{y}_0=m_0,l_0,b_0,r_0,\mu_{\alpha^\ast,0},\mu_{\delta 0},v_{r 0})$ is the vector describing the `true' underlying stellar properties unaffected by observational errors.

We can then define the un-normalised\footnote{Although the PDF must be normalised, it is convenient to define the un-normalised PDF first, and then come back to the normalisation constant when all of the components of the PDF have been defined.} PDF such that
\begin{equation}\mathcal{D}(\mathbf{y}_i|\boldsymbol{\theta})=\mathcal{S}(\mathbf{y}_i) \int_{\forall \mathbf{y}_0} \varphi_{M_0} \varphi_{rlb_0}   \varphi_{v_0}  \mathcal{E} (\mathbf{y}_i|\mathbf{y}_0) \, d\mathbf{y}_0
,\end{equation}
where $\varphi_{M_0}$, $\varphi_{rlb}$,  $\varphi_{v}$ are the PDFs describing the nature of the open cluster, and $\mathcal{S}(\mathbf{y})$ is the selection function, which takes the probability of observing a star into account, given the properties of the star and the instruments' observational capabilities. To take the fact that Hipparcos is a magnitude limited sample into account, a step function is used with 
\begin{equation} 
\mathcal{S}(\mathbf{y}) =\begin{cases}
    1, & \textrm{if $H_p$\textless 12.5}.\\
    0, & \textrm{otherwise}.
  \end{cases} 
  \label{eqn:HipLim}
  \end{equation}

The case is more complicatedin reality , with Hipparcos only being complete up to magnitude $H_p<7$. At fainter magnitudes, stars were selected using a number of criteria and used as an input catalogue \citep{HipparcosInput}. Hipparcos had a physical limit in the number of stars observable in a single field of view, and it suffered from glare effects on the telescope when observing stars very close together on the sky. Therefore, decisions were made on a case-by-case basis as to which stars to observe, depending on the number and position of stars in each field of view. The Hipparcos Mission Pre Launch Status Report states that: ``\emph{Once the list of likely cluster members had been established and the worst veiling glare cases excluded, a somewhat arbitrary compromise had to be found}''. The hand selection of stars for observation makes it impossible to accurately model a selection function based on apparent magnitude that can describe the selection probability for the underlying cluster population, which does not strictly follow the definition given in Eq. \ref{eqn:HipLim}. However, as the Hipparcos star selection mostly depended on the proximity of target stars to each other on the sky and not on magnitude, the effects on the results given in Sects. \ref{sec:pleiades} and \ref{sec:hyades} are minimal.

\subsection{Models}
The distribution of absolute magnitudes in a given photometric band is assumed to be a Gaussian distribution around a mean colour-absolute magnitude relation, $M_{\rm mean}$, with some dispersion, $\sigma_M$:
\begin{equation}
\varphi_{M} = e^{-0.5 \left( \frac{M-M_{\rm mean}(V-I)}{\sigma_{M} }\right)^2}
\label{eqn:M}
.\end{equation} 
Here, the absolute magnitude is given by
\begin{equation}
M = m + 5 \textrm{log}_{10} (\varpi) +5 - A
\label{Mmrelation}
,\end{equation} 
where $m$ is the apparent magnitude of the star, $\varpi$ its parallax, and $A$ is the interstellar extinction of the star in the same photometric band. 

In the following work, the interstellar extinction is assumed to be known, although it can be left as a free fit parameter. For the Pleiades, a single extinction value of $A_{H_p}$=0.0975 magnitudes in the Hipparcos band H is used for all members. This is derived from a reddening of $E(B - V) = 0.025 \pm 0.003$ found by \cite{Extinction}, which is converted to an extinction estimate in the visual band through $A_V = 3.1 E(B-V)$ and then into the Hipparcos band $H_p$. For the Hyades, the level of extinction is assumed to be negligible, and an extinction of $A_{H_p}$=0 is assumed. If available, individual extinction estimates for each star could be used, in order to correctly take effects of differential reddening into account.

The term $M_{\rm mean}(V-I)$ gives the mean absolute magnitude of stars as a function of colour, while $\sigma_{M}$ is the intrinsic dispersion. For simplicity, known binaries are removed. Additionally, the given distribution does not support the giant branch, since this would require the magnitude function to turn back on itself, giving a non-unique solution. Therefore, giants are also removed, enabling the application of this method to all clusters, irrespective of age. 

The fitting procedure has two options: 
\begin{enumerate}
  \item It fits the position of points on the colour-absolute magnitude diagram, to which a spline function is fitted in order to determine a colour-dependent absolute magnitude distribution approximating the isochrone of the cluster; or  \item  If a theoretical isochrone is supplied as input, the shape of the magnitude distribution is taken from the isochrone. While the shape of the isochrone is conserved, the isochrone is free to be shifted in absolute magnitude. 
\end{enumerate}

If using the second option, the age and metallicity of the cluster must be assumed. This makes the first option preferable for clusters with little information on age and metallicity, because parameters can be determined without having to be concerned with models that depend on this information.

The spatial distribution we have assumed follows a spherical Gaussian distribution in Cartesian coordinates that, expressed in a rotated galactic coordinate system (see Appendix \ref{Sec:CoordinateConversion}), follows:
\begin{equation} 
 \varphi_{rl'b'} = r^2 \textrm{cos}(b')  e^{ - \frac{0.5}{\sigma_S^2} \left(R^2 + r^2 -2rR \textrm{cos}(b') \textrm{cos}(l') \right)} 
\end{equation}
where the term $r^2 \textrm{cos}(b')$ is the Jacobian of the coordinate transformation, $R$  the mean distance to the cluster, $r$  the distance to the individual star, and $l'$ and $b'$ are the rotated coordinates of the star.

The Gaussian spatial distribution was chosen for ease of implementation, and as a first approximation of the cluster's spatial structure. It is possible to use a distribution specifically suited to open clusters, such as King's profile \citep{KingProfile}, which could make the distribution more realistic. This is a possible improvement for later additions to this work.  

Finally, the velocity distribution in Cartesian coordinates is defined as the velocity ellipsoid:
\begin{equation} 
\varphi_{v} = e^{-0.5 \left( \frac{U-U_{\rm mean}}{\sigma_{U} }\right)^2 -0.5 \left( \frac{V-V_{\rm mean}}{\sigma_{V} }\right)^2 -0.5 \left( \frac{W-W_{\rm mean}}{\sigma_{W} }\right)^2}
,\end{equation}
where $U_{\rm mean}$, $V_{\rm mean}$, and $W_{\rm mean}$ are the components of the cluster's mean velocity along each Cartesian axis.

The distribution in observational errors is given by 
%\begin{equation} \mathcal{E} (y|y_0)= 
%e^{-0.5 \left( \frac{\varpi-\varpi_0}{\epsilon_\varpi}\right) ^2 } 
%e^{-0.5 \left( \frac{\mu_{\alpha^\ast}-\mu_{\alpha^\ast,0}}{\epsilon_{\mu_{\alpha^\ast}}}\right) ^2 } 
%e^{-0.5 \left( \frac{\mu_{\delta}-\mu_{\delta,0}}{\epsilon_{\mu_{\delta}}}\right) ^2 } 
%e^{-0.5 \left( \frac{v_r-v_{r0}}{\epsilon_{v_r}}\right) ^2 } \delta(m,l',b') \end{equation}

\begin{equation} \mathcal{E} (\mathbf{y}|\mathbf{y}_0)= \mathcal{E} (\varpi|\varpi_o) \mathcal{E} (\mu_{\alpha^\ast}|\mu_{\alpha^\ast,0}) \mathcal{E} (\mu_{\delta}|\mu_{\delta,0})\mathcal{E} (v_r|v_{r_0})  \delta(m,l',b').\end{equation}
All observational errors are assumed to follow a Gaussian distribution with a variance given by the formal error $\epsilon_\mathbf{y}$. Here, $\varpi$ is the parallax, $\mu_{\alpha^\ast}$ and $\mu_\delta$ the proper motions, and $m$ the apparent magnitude.
The delta function $\delta(m,l',b')$ describes the case for which a parameter's observational error is small enough to be deemed negligible.

An additional benefit of using this formulation with `true' object parameters in the models and then linking these true parameters with the observed quantities is that observational data including negative parallaxes can be used directly without problems attempting to calculate the logarithm of a negative number (e.g. in Eq. \ref{Mmrelation}). The inclusion of stars with negative parallaxes is essential for avoiding biasing the sample by preferentially removing more distant stars and biasing
the average distance by selecting the stars with only positive errors.

The normalisation constant required for $ \mathcal{O}(\mathbf{y}_i|\mathbf{\theta}) $ is found by integrating the un-normalised joint probability distribution $\mathcal{D}(\mathbf{y}|\mathbf{\theta})$ over all $\mathbf{y}$:

\begin{equation} 
\mathcal{C} = \int_{\forall \mathbf{y}_0} \int_{\forall \mathbf{y}}\varphi_{M_{0}}\varphi_{\varpi_0l_0'b_0'}  \varphi_{v} \mathcal{S}(\mathbf{y}) \mathcal{E}(\mathbf{y}|\mathbf{y}_0) \, d\mathbf{y} \, d\mathbf{y}_0
.\end{equation}
The exact analytical solution has been found where possible, and the remaining integrals with no analytical solution are solved numerically (See Appendix \ref{sec:IntegrationOfLikelihoodFunction}). 

\subsection{Formal errors}

The Hessian matrix is constructed through numerical differentiation of the likelihood function at its global maximum. The inverse of the Hessian matrix is the covariance matrix, and the formal errors are calculated from the square root of the diagonal of the covariance matrix. After calculating the covariance matrix and formal errors, correlations between each of the parameters can easily be obtained. 

\subsection{Data binning}

While some parameters, such as mean distance, are `global' parameters describing some general property of the open cluster, a number of the parameters show a dependence on colour. For example, the physical spatial distribution of some clusters is believed to change as a function of mass, hence colour, through the process of mass segregation. 

With a sufficiently precise data set containing enough stars, it is possible to fit a smooth function describing a parameter's colour dependence, if applicable, by estimating the parameters of, for example, a polynomial approximation of the dependence. Owing to the limited available data in the Hipparcos catalogue for the Pleiades and Hyades, there is insufficient information to constrain such distributions in all cases, so where necessary an approximation has been made by binning the data. 

A star's normalised PDF can be thought of as the sum of several other PDFs, such that
\begin{equation}  
\frac{\mathcal{D}(\mathbf{y}_i | \boldsymbol{\theta})}{\mathcal{C}} = \frac{  \mathcal{W}(V-I)_1 \mathcal{D}(\mathbf{y}_i | \boldsymbol{\theta}) + \mathcal{W}(V-I)_2 \mathcal{D}(\mathbf{y}_i | \boldsymbol{\theta})  + ...}{\mathcal{C}} 
\end{equation}  

where $\mathcal{C}$ is the normalisation constant and $\mathcal{W}$(V-I) a selection function dependent on colour:
\begin{equation} 
\mathcal{W} =\begin{cases}
    w, & \textrm{if $ (V-I)_{min}<(V-I)_{star}<(V-I)_{max}$}.\\
    0, & \textrm{otherwise},
  \end{cases} 
  \end{equation}
where $w$ is a normalised coefficient that depends on the relative number of stars per colour bin.   

The normalisation constant $C$ is found by integrating the PDF over all $\mathbf{y}$:

\begin{equation} 
\mathcal{C}= \int_{\forall \mathbf{y}} \mathcal{D}(\mathbf{y}|\boldsymbol{\theta}) \,\,d\mathbf{y} 
\end{equation}
\begin{equation} 
\equiv \int_{\forall \mathbf{y}} \mathcal{W}(V-I)_1 \mathcal{D}(\mathbf{y}|\boldsymbol{\theta}) d\mathbf{y} + \int_{\forall \mathbf{y}} \mathcal{W}(V-I)_2 \mathcal{D}(\mathbf{y}|\boldsymbol{\theta}) d\mathbf{y} +  ... 
\end{equation}

The strength of this approach is that parameters with a single value of interest (with no colour dependence) are described by only one parameter in $\boldsymbol{\theta}$. Where more information can be gained by producing an estimate of a parameter in each colour bin, it is possible to define a separate parameter for each colour bin. 

The parameters to be estimated therefore become

\begin{equation}
 \boldsymbol{\theta} = (R,\sigma_R^{n},M^{n+1},\sigma_M^{n},U,V,W,\sigma_{UVW})
 ,\end{equation}
where $n$ is the number of bins, $R$  the mean distance to the cluster (assuming a spherical Gaussian distribution), $\sigma_R^n$  the intrinsic dispersion around the mean distance in each colour bin, $M^{n+1}$ are the absolute magnitude values used for fitting the spline function of the isochrone, $U$, $V$, and $W$ are the three components of the velocity ellipsoid, and $\sigma_{UVW}$ is the velocity dispersion.

\subsection{Testing with simulations}

The method described in Sect. \ref{sec:math} was implemented and tested extensively using simulations. During development and testing of the MLE implementation, simulated catalogues were constructed using Monte Carlo techniques. Each star in the simulated population is given a position in Cartesian coordinates, which is then converted into a sky position and distance, and given a velocity in Cartesian coordinates, which is converted into proper motions and a radial velocity. The values are randomly chosen from a spherical Gaussian spatial distribution and a velocity ellipsoid, with the mean and variance of each distribution chosen by hand. The relationship between colour and absolute magnitude is assumed to be linear for simplicity.

Basic simulation of observational errors was achieved by adding an error value derived from a Gaussian random number generator to each parameter
in the simulated catalogue, with the variance chosen to be a value similar to the errors found in the Hipparcos catalogue.
Additionally, the more realistic AMUSE open cluster simulator \citep{AMUSE} was used to simulate open clusters with more realistic distributions, including a realistic isochone.

\begin{figure}
\begin{center}
\includegraphics[width=\linewidth]{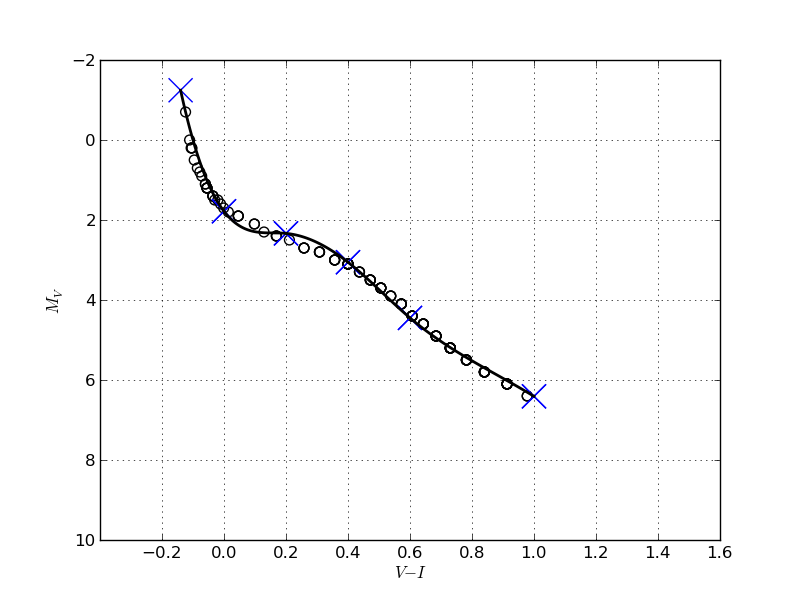} 
\caption{CMD for a simulated Pleiades like cluster found in the GaiaSimu library, with a distance of 2.6 kpc. Open circles show the `true' simulated absolute magnitudes without errors. Blue crosses are the points fitted with the ML method applied to the data after the simulation of observational errors. The solid line is the resulting isochrone-like sequence, showing accurate reconstruction of the isochrone-like sequence from error effected data. }
\label{fig:testingWithSims}
\end{center}
\end{figure}

The final stage of testing with simulations uses a set of 500 simulated open clusters found in the Gaia Simulator (\cite{masana}, \cite{GUMS}) in order to test the suitability of the method in use with Gaia data (see Sect. \ref{sec:OutlookForGaia}). The Gaia Object Generator \citep{GOG} was used to simulate realistic errors as expected for Gaia catalogue data. Using the ML method on this simulated data showed the successful extraction of an isochrone-like sequence from error-effected data (Fig. \ref{fig:testingWithSims}) and could reproduce the input distance of the cluster, within formal error bounds and, after repeated testing, with no significant bias.

In the following sections (\ref{sec:data}, \ref{sec:pleiades}, \ref{sec:correlations}, and \ref{sec:hyades}), we apply the method to the best currently available data on the Pleiades and the Hyades. In Sect. \ref{sec:OutlookForGaia} we use simulations to extrapolate the use of the method out to greater distances, in expectation of the Gaia data.

\section{Data}
\label{sec:data}
The method described here has been applied to the new Hipparcos reduction (\cite{newhipcat}) data for 54 well known Pleiades members. The 54 cluster members are believed to be a clean sample and have been identified and used in numerous previous studies (e.g. \cite{FVL1997}, \cite{robichon} and \cite{Makarov}). 

The new Hipparcos reduction has several advantages over the original Hipparcos catalogue, including a reduction in the formal errors by up to a factor of 4 for the brightest stars and a claimed reduction, by up to a factor of 10, in the correlations between stars observed over small angles. This was achieved through a completely new method for formulating the Hipparcos satellite's attitude model, replacing the previous `great-circles' reduction process with a fully iterative global solution. This new method of catalogue reconstruction supersedes an earlier attempt by \cite{Makarov} to reduce correlation in the Hipparcos catalogue specifically for the Pleiades open cluster.

This reduction in correlations is particularly important in the study of open clusters. This is because correlated errors in the original Hipparcos catalogue were blamed for the cases of discrepancy between distances calculated with Hipparcos parallaxes, compared with photometric data or other methods.

Where radial velocity information has been used, it was obtained from (\cite{RV1}, \cite{RV2}, \cite{RV3}, \cite{RV4}) through the WEBDA database.

\section{Results - Pleiades}
\label{sec:pleiades}
The results of the application of the method described in Sect. \ref{sec:math} can be seen in Tables \ref{table:results1} and \ref{table:results2}. Parameters describing the general properties of the cluster, such as mean distance and mean proper motion are given in Table \ref{table:results1}, whereas colour-dependent parameters are given in Table \ref{table:results2}. For the latter, stars were divided into four color bins. The choice of bins is arbitrary and have been selected to give roughly the same number of stars per bin. 

\subsection{Distance}
The mean distance to the Pleiades has been estimated to be $120.3 \pm 1.5$ pc. This agrees with \cite{FVL20Clusters}, who finds $120.2\pm2.0$ pc using the same dataset. The formal errors assigned to each parameter are calculated from the square root of the covariance matrix of the likelihood function. The formal error is approximately 25\% smaller than from \cite{FVL20Clusters}, with the increased precision from the use of this method attributed to the inclusion of parallax, position, proper motion, colour, and apparent magnitude information.

The distance is given as a general property of the cluster alone (Table \ref{table:results1}), because there is no physical reason to expect differing distances for different colour bins, unlike, for example, the size parameter, $\sigma_R$.

The mean distance to the cluster has been included as a colour-dependent parameter during testing, in order to check the ML method functions as expected and to check that there are no biases present in the Hipparcos catalogue. The mean distance to the cluster was found to be consistent across all colour bins, within the error bounds. This is a good test that there is no colour- dependent bias in the Hipparcos Pleiades data.

\subsection{Kinematics}
Testing was carried out for two cases: first where only Hipparcos data has been used, and second including the use of radial velocity data where available. Fifteen Hipparcos Pleiades members have radial velocity data (less than one third of the sample). Working in a mixed case, where some stars have radial velocity information and some do not, is possible through marginalisation of terms with missing data from the joint PDFs. 

With the inclusion of radial velocity data, the variance in the space velocity has been found to be nearly $6$ kms$^{-1}$. We believe that a lack of homogeneity in the data is responsible for the large variance, because the data is compiled from several sources. While there is no change in the estimation of the mean distance to the cluster between the two cases, the formal errors are in fact larger with the inclusion of radial velocity data. Therefore, the lack of an accurate and homogeneous catalogue of radial velocities for a significant number of Hipparcos Pleiades stars means that radial velocity information has been ignored, and the results shown are for a fitting to a Gaussian distribution in proper motion rather than a three-dimensional space velocity. 

The variance in the distribution in proper motions is found to be $1.7\pm0.3$ and $1.5\pm0.2$ mas year$^{-1}$ for $\mu_{\alpha^\ast}$ and $\mu_\delta$, respectively. Taking the distance to be 120.3 pc, this variance is equivalent to a variance in the velocity distribution of the cluster of $1.3\pm0.4$ kms$^{-1}$.

\subsection{Size}
   
The spatial distribution of the open cluster is modelled using a spherical Gaussian distribution, with its center at some distance $R$ and a variance around the mean distance $\sigma_R$. The variance has been included as a colour-dependent term and is estimated for each colour bin. The results show an increase in $\sigma_R$ with an increase in colour $(V-I)$, which corresponds directly to a decrease in stellar mass.    
   
As the variance in the spatial distribution increases from 3.3 pc to 13.2 pc over the full colour range for the observed stars, we find strong evidence for mass segregation. This relationship between stellar mass and the spatial distribution of stars in a cluster has been reported for the Pleiades with a similar degree of segregation, using a star counting technique for the two-dimensional case of the cluster projected onto the plane of the sky \citep{pleiadesMassSegregation}.

\subsection{Absolute magnitude distribution}
\label{sec:PleiadesMag}
An approximation to the isochrone of the cluster is found by fitting points in the colour-absolute magnitude diagram to find the absolute magnitude distribution of the star as a function of colour. A spline function converts these points into a smooth line, which can be thought of as the observed isochrone of the cluster. The points used for the spline function are found by the fitting method and labelled A and B for each colour bin in Table \ref{table:results2}. The resolution of the resulting fit only depends on the number of stars, which limits the possible number of bins, and the precision of the data. 

The intrinsic dispersion, $\sigma_M$, is the dispersion in absolute magnitude around the mean magnitude given by the magnitude distribution, over any narrow colour range $(V-I)$ to $(V-I)+\delta_{(V-I)}$. Since the dispersion is not constant over the whole colour range, the value of $\sigma_M$ is given for each colour bin. 

\begin{figure}
\begin{center}
\includegraphics[width=\columnwidth]{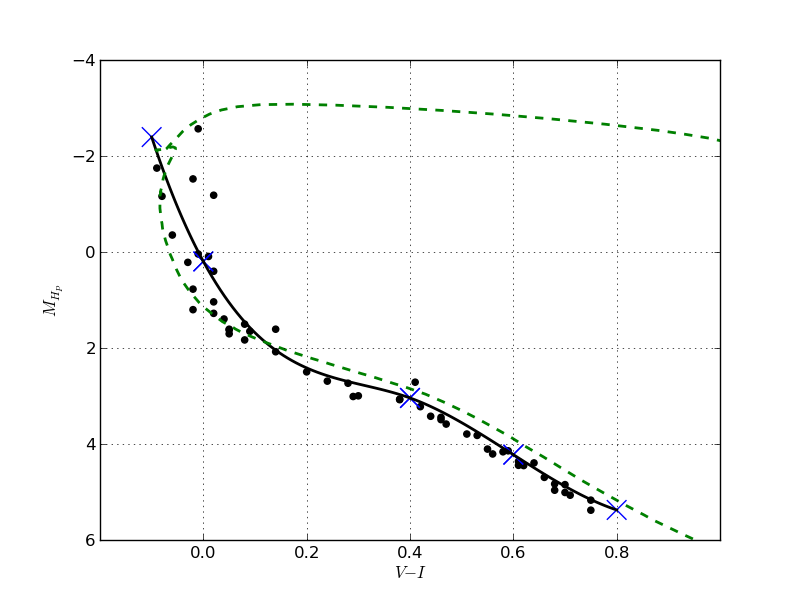} 
\caption{Colour-absolute magnitude diagram for the 54 Hipparcos Pleiades members. $M_H$ is the absolute magnitude in the Hipparcos photometric band calculated using the posterior parallax. The blue crosses are the results of the fitting, with the spline function giving the magnitude dependence as the thick line. The green dashed line is the theoretical isochrone generated using PARSEC v1.1, with an age of 100 Myr, and $Z$ = 0.03 \citep{bressan}. }
\label{fig:CMDpleiades}
\end{center}
\end{figure}

The results of the fitting can be seen in Table \ref{table:results2}, where they have been plotted onto the colour magnitude diagram of the Pleiades in Fig. \ref{fig:CMDpleiades}. The theoretical isochrone taken from the PARSEC library \citep{bressan} can be seen in green. The isochrone was generated using PARSEC v1.1, with an age of 100 Myr, and $Z$ = 0.03. 

Excluding the turn off ($V-I<$ 0.1), the shape of the theoretical isochrone is accurately obtained by the fitting procedure (see Fig. \ref{fig:CMDpleiades}). The {\raise.17ex\hbox{$\scriptstyle\sim$}}0.3 mag difference in absolute magnitude between the theoretical isochrone and the sequence found by the fitting procedure is the discrepancy historically reported between Hipparcos and photometry-based methods.

Above the main sequence turn off, neither the theoretical isochrone nor the results from this work accurately fit the data. This is due to the presence of stars migrating to the giant branch. 

The intrinsic dispersion of absolute magnitude around the isochrone is found for each colour bin. The large dispersion in the first bin is due to the presence of the turn off, and subsequently the dispersion around the main sequence decreases with increased $V-I$.

\begin{table}\centering
   \begin{tabular}{@{}rrrcrrcrrcrr@{}}\toprule  
        Parameter       & Estimated   & Error \\ \midrule
    Distance (pc)     & 120.3       &  1.5      \\ 
    $\mu_{\alpha^\ast}$ (arcsec year$^{-1}$) &   19.9  &  0.3     \\
    $\mu_\delta$ (arcsec year$^{-1}$) &   -45.3  & 0.3     \\
    $\sigma_{\mu_{\alpha^\ast}}$ (arcsec year$^{-1}$) &  1.7  & 0.3 \\          
    $\sigma_{\mu_\delta}$ (arcsec year$^{-1}$) &  1.5  &  0.2 \\          
    \bottomrule
    \end{tabular}
    \caption{Colour-independent results obtained for the Pleiades from the method applied to the new Hipparcos reduction.}
    \label{table:results1}
\end{table}
 
\begin{table*}\centering
   \begin{tabular}{@{}rrrcrrcrrcrr@{}}\toprule  
    &  \multicolumn{2}{c}{ $(-0.1<V-I<0.0)$ }  & \phantom{} & \multicolumn{2}{c}{ $(0.0<V-I<0.4)$ } & \phantom{} & \multicolumn{2}{c}{$(0.4<V-I<0.6)$} & \phantom{} & \multicolumn{2}{c}{$(0.6<V-I<0.8)$}\\  
           \cmidrule{2-3} \cmidrule{5-6} \cmidrule{8-9}  \cmidrule{11-12}
        Parameter       & Estimated    & Error & & Estimated    & Error & & Estimated    & Error & & Estimated    & Error\\ \midrule
    $\sigma_R$ (pc)   &  3.4  &  1.0    && 4.9   &  0.9   && 10.3   & 2.9   && 13.1  & 3.8  \\ 
    $\sigma_M$ (mag)  &  1.6  & 0.5     && 0.45  &  0.06  &&  0.22  & 0.06  && 0.17  & 0.06   \\
    A (start point)   & -2.4  & 0.7     && 0.2   &  0.2   &&  3.0   &  0.1  && 4.2   &0.1  \\
    B (end point)     & 0.2   & 0.2     &&  3.0  &  0.1   &&  4.2   &  0.1  && 5.4   & 0.3  \\
    \bottomrule
    \end{tabular}
    \caption{Colour-dependent results obtained for the Pleiades from the method applied to the new Hipparcos reduction. In the four bins there are 9, 21, 12, and 12 stars (low $(V-I)$ to high). A and B are the points found for the spline function fitting to the colour-absolute magnitude relationship, with A at the start of the bin and B at the end. Each corresponds to a cross in Fig. \ref{fig:CMDpleiades}.}\label{table:results2}
\end{table*}

\begin{table}\centering
   \begin{tabular}{@{}rrrcrrcrrcrr@{}}\toprule  
        Parameter       & Estimated   & Error \\ \midrule
    Distance (pc)     & 122.1       &  3.7      \\ 
    $\mu_{\alpha^\ast}$ (arcsec year$^{-1}$) &   20.0  &  0.5     \\
    $\mu_\delta$ (arcsec year$^{-1}$) &   -45.9  & 0.3     \\
    $\sigma_{\mu_{\alpha^\ast}}$ (arcsec year$^{-1}$) &  2.0  &  0.4 \\          
    $\sigma_{\mu_{\delta}}$ (arcsec year$^{-1}$) &  1.3  &  0.3 \\    

    \bottomrule
    \end{tabular}
    \caption{Results obtained for the Pleiades after cutting all stars at or above the apparent location of the main sequence turn off: $(V-I)>0.1$}
\end{table}

\begin{table*}\centering
   \begin{tabular}{@{}rrrcrrcrr@{}}\toprule  
    &  \multicolumn{2}{c}{ $(0.1<V-I<0.4)$ } & \phantom{} & \multicolumn{2}{c}{$(0.4<V-I<0.6)$} & \phantom{} & \multicolumn{2}{c}{$(0.6<V-I<0.8)$}\\  
           \cmidrule{2-3} \cmidrule{5-6} \cmidrule{8-9} 
        Parameter       & Estimated    & Error & & Estimated    & Error & & Estimated    & Error\\ \midrule
    $\sigma_R$ (pc)   &   9.3   &  2.5   && 8.2    & 2.3   && 12.5   & 3.6  \\ 
    $\sigma_M$ (mag)      &   0.14   &  0.06   &&  0.22   & 0.06   && 0.13   & 0.06   \\
    A (start point)     & 1.6  &   0.2  && 3.1  &  0.1      && 4.1  &0.1  \\
    B (end point)       &   3.1  &  0.1   &&  4.1   &  0.1  && 5.8   & 0.2  \\
    \bottomrule
    \end{tabular}
    \caption{Results obtained for the Pleiades after cutting all stars at or above the apparent location of the main sequence turn off. The results remain unchanged within the error range. In the three bins there are 9, 12, and 12 stars (low $(V-I)$ to high).}
\end{table*}

\section{Correlations}
\label{sec:correlations}
The main argument against Hipparcos-based distance estimates of open clusters has been that Hipparcos trigonometric parallaxes have correlated errors on small angular scales. Indeed, this has been cited as the cause of the large (0.3 mag) discrepancy in the Pleiades distance between Hipparcos-based and photometry-based methods. \cite{HipCorrelatedErrors} argue that correlated parallax errors cause a stark difference between Hipparcos trigonometric parallaxes and the Hipparcos kinematic parallaxes derived from proper motions.

According to \cite{HipCorrelatedErrors}, the proper-motion-based parallax can be determined from Hipparcos data using
\begin{equation}  
\varpi_{\textrm{pm},i} = \frac{ \Braket{(\boldsymbol{V}_t)_i |  \boldsymbol{C}_i^{-1}  |  \boldsymbol{\mu}_{\textrm{HIP},i} } }{  \Braket{(\boldsymbol{V}_t)_i |  \boldsymbol{C}_i^{-1}  | (\boldsymbol{V}_t)_i  } }
\label{nandg}
,\end{equation}
where $(\boldsymbol{V}_t)_i$ is the transverse velocity of the cluster in the plane of the sky at the position of the star $i$,  $\boldsymbol{C}_i$ is the sum of the velocity dispersion tensor divided by the square of the mean distance to the cluster and the covariance matrix of the Hipparcos proper motion, and $ \boldsymbol{\mu}_{\textrm{HIP},i}$ is the vector describing the proper motion of the star.

\begin{figure}
\begin{center}
\includegraphics[width=\columnwidth]{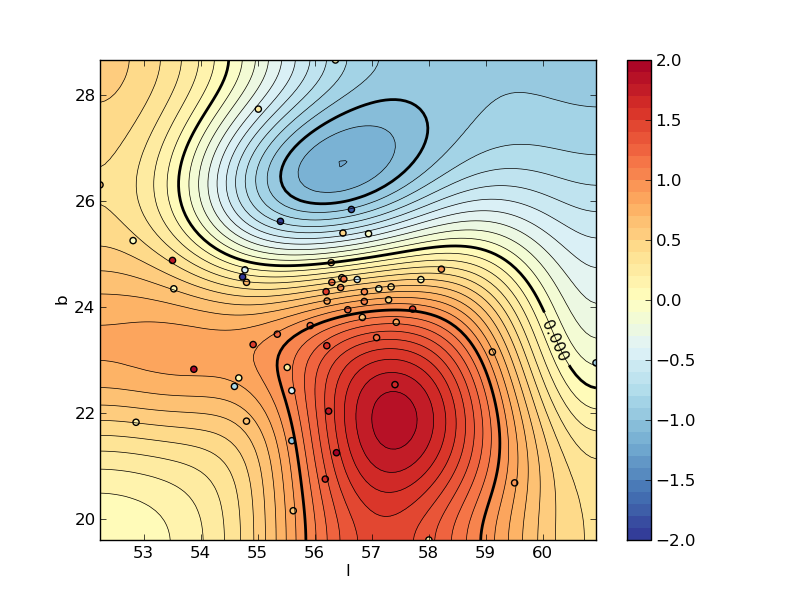} 
\caption{Smoothed contours of the difference between the Hipparcos parallaxes and the proper-motion-based parallax ($\varpi_{Hip}-\varpi_{pm}$) of the 54 Pleiades cluster members, with data taken from the original Hipparcos catalogue (1997). Thin contours are at intervals of 0.1 milli-arcseconds, thick contours at intervals of 1 milli-arcsecond.}
\label{fig:contourOld}
\includegraphics[width=\columnwidth]{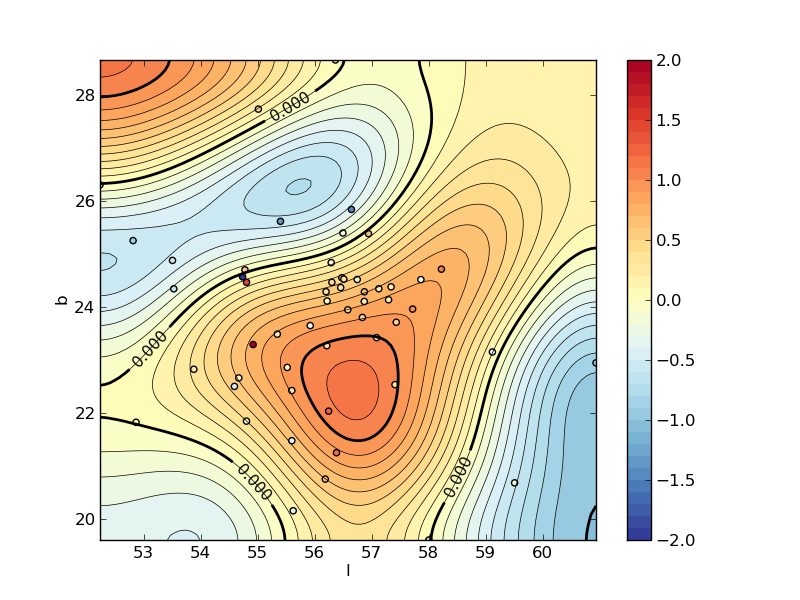} 
\caption{Same as Fig. \ref{fig:contourOld}, but with parallax data taken from the new Hipparcos reduction (2007).}.
\label{fig:contourNew}
\includegraphics[width=\columnwidth]{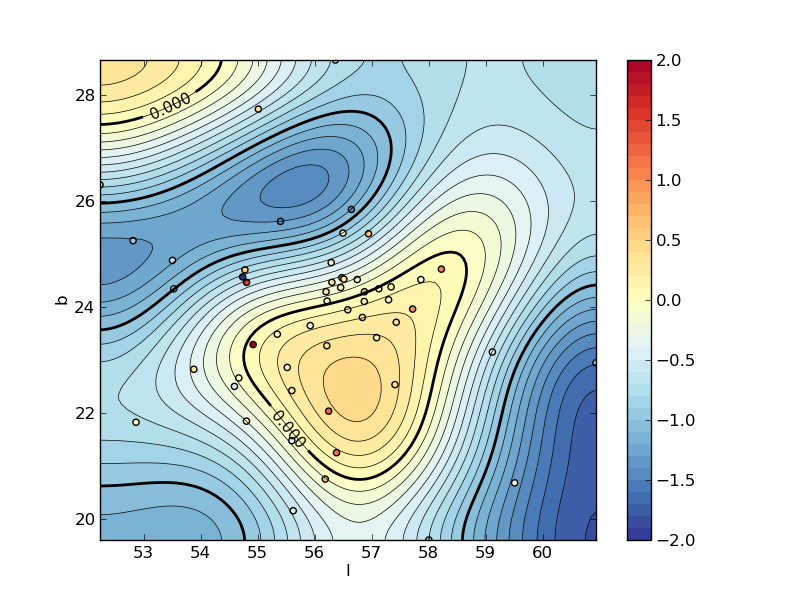} 
\caption{Same as Fig. \ref{fig:contourNew}, but with an assumed mean cluster distance of 120 pc as implied by the Hipparcos data for the computation of the proper motion based parallax.}
\label{fig:contourNew120}
\end{center}
\end{figure}

Narayanan and Gould used a plot showing the contours of the difference between Hipparcos parallaxes and the parallaxes derived from proper motions (Eq. \ref{nandg}) to argue that the Hipparcos parallaxes are systematically larger by up to two milli-arcseconds throughout the inner $6\,^{\circ}$ of the Pleiades. This figure has been recreated in Fig. \ref{fig:contourOld} with the 54 member stars used in this work.

Figure \ref{fig:contourNew} has been produced using the method described by \cite{HipCorrelatedErrors}, but using parallax data from the new Hipparcos reduction. While in the original plot from Narayanan and Gould there is clearly a region where the trigonometric parallaxes are larger than those derived from proper motions, this feature has been reduced in severity by a factor of two by using the new Hipparcos reduction. 

This new figure confirms, for the case of the Pleiades, claims made by \cite{newhipcat} that correlations in the new reduction have been reduced significantly. Additionally, this disagrees with claims that the shorter distance for the Pleiades derived from Hipparcos is due to the correlated errors in parallaxes for Pleiades stars. If this was the case, one would expect the distance estimate from the new Hipparcos reduction to be greater now that the correlations have been reduced. 

In fact, the distance derived from the new reduction in this work and by \cite{FVL20Clusters} put the Hipparcos distance to the Pleiades at $2.5$ pc longer than those derived from the original Hipparcos catalogue, a much smaller difference than the roughly 10\% historic discrepancy. The method presented here has also been applied to the original Hipparcos catalogue. The difference between the estimated distance from the old and new Hipparcos reductions is only 2\%. The small change in distance estimate despite the reduction in correlations by a factor of 10 implies that correlations cannot be responsible for the long-standing discrepancy.

Additionally, in the calculation of the proper-motion-based parallax, a mean distance to the cluster must be assumed. \cite{HipCorrelatedErrors} derived a distance to the Pleiades of 131 pc using Hipparcos proper motions, which was then used in calculating the proper-motion-based parallaxes that form the basis of Fig. \ref{fig:contourOld} and their argument against Hipparcos. Using the distance of 120 pc as found in this work and implied by studies using Hipparcos parallaxes, the baseline for the correlation plots is shifted, as can be seen in Fig. \ref{fig:contourNew120}. In this final figure no significant residual biasing the results is apparent, contrary to the original claims.

\section{Results - Hyades}
\label{sec:hyades}
This method has also been applied to the new Hipparcos reduction for the Hyades open cluster. Of the 282 potential Hyades members used by \cite{Perryman1998}, a detailed study by \cite{JDBhyades} finds 218 probable members, which are used as the basis of this study. Radial velocities from ground-based observations have been collected in \cite{Perryman1998} and used extensively here. 

The star HIP20205 was rejected as a giant, because the isochrone fitting does not currently support fitting of the giant branch.

\cite{Galli} reject the star HIP28774 in their analysis of the Hyades cluster due to conflicting data for this star in the old and new Hipparcos reductions. In the following sections, this star is not present in our membership list after selecting only the stars in the inner 10 pc. When using the full 218 probable members, the star is included, however its removal has no real effect on the results. This is due to the very low precision of the data on this star in the Hipparcos catalogue, and therefore its very low statistical weight within the method.

\subsection{Distance}
As in \cite{Perryman1998}, we select only the stars within the inner 10 pc for the distance estimation, since the large spatial dispersion and presence of numerous halo stars is not modelled well by a spherical Gaussian distribution. The distance to the Hyades has been estimated as $46.35 \pm 0.35$ pc. This is slightly more than \cite{Perryman1998}, who finds 46.34 $\pm$ 0.27 pc, and slightly smaller than \cite{FVL20Clusters}, who finds 46.45 $\pm$ 0.50 pc. \cite{Perryman1998} used the original Hipparcos catalogue, whereas \cite{FVL20Clusters} used the new reduction from 2007. Both authors used differing membership selection and determination techniques.

The three-dimensional dispersion in the spatial distribution of the cluster's core is estimated to be 3.4 $\pm$ 0.2 pc, which is consistent with the observed distribution in sky position. Assuming this dispersion in the spatial distribution, the selection of stars with cluster radius of less than 10 pc corresponds to a 3 $\sigma$ limit, and so the spatial distribution of the core of the cluster should not be significantly affected.

Using the ML method on the full 218 probable member stars, including halo stars, the mean distance is found to be 42.6 $\pm$ 0.5 pc. The decrease in precision with the increase in the number of stars is attributed to the clearly non-Gaussian distribution of a dense core and a large number of disperse halo stars. In this case the spatial distribution of the cluster's core is estimated to be 15.8 $\pm$ 3.9 pc. The modelling of the spatial distribution could be improved through the use of a King's profile \citep{KingProfile} or an exponential distribution. This is being considered for future improvements.

That member stars are found with distances from the center greater than its tidal radius of {\raise.17ex\hbox{$\scriptstyle\sim$}}10 pc \citep{madsen} is reasonable, and is expected due to the large number of so-called halo stars. These stars have been found to exist in the Hyades at a radius of 10 to 20 pc \citep{brown}, and although they are effected by the galactic gravitational field, they remain bound to the cluster for some significant time.

\subsection{Kinematics}
\label{sec:hyadesKinematics}

The space velocity of the Hyades has been found to be $46.5\pm 0.2 $ kms$^{-1}$, with an internal dispersion on the velocity of $1.11 \pm 0.05$ kms$^{-1}$. The internal dispersion is slightly larger than those in the literature reviewed by \cite{JDBhyades}, who find a dispersion of {\raise.17ex\hbox{$\scriptstyle\sim$}}0.3 kms$^{-1}$. 

As for the case of the Pleiades, the high velocity dispersion is attributed to an inhomogeneous radial velocity data. As highlighted by \cite{brown}, the radial velocity data for the Hyades comes from a combination of sources with greatly varying precision and zero points. 

\subsection{Absolute magnitude distribution}

\begin{figure}
\begin{center}
\includegraphics[width=\columnwidth]{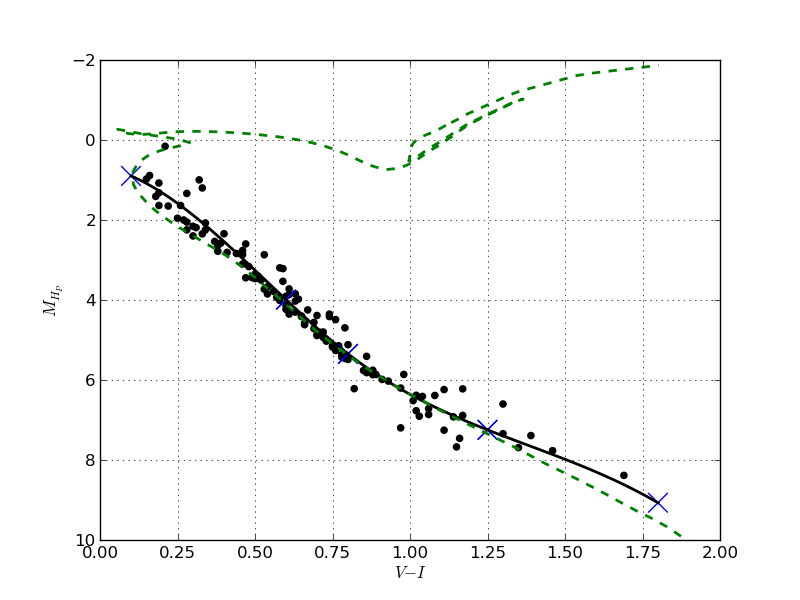} 
\caption{Colour-absolute magnitude diagram for the 128 Hipparcos Hyades members found in the inner 10 pc of the clusters core. $M_{H_P}$ is the absolute magnitude in the Hipparcos band, and is calculated using the posterior parallax. The blue crosses are the results of the fitting, with the spline function giving the magnitude dependence as the thick line. The dashed line is the theoretical isochrone from the PARSEC library, with an age of 630Myr and $Z = 0.024$.}
\label{fig:CMDhyades}
\end{center}
\end{figure}

As described in Sect. \ref{sec:PleiadesMag}, an estimate of the isochrone of the cluster is produced through fitting a smoothed line to the mean absolute magnitude $M_{\rm mean}$ in Eq. \ref{eqn:M}. The results of the fitting can be seen in Fig. \ref{fig:CMDhyades}, with the theoretical isochrone overlaid in green.

In contrast to the results for the Pleiades, the isochrone from the ML method and the theoretical isochrone from the PARSEC library are in strong agreement in both shape and position over most of the main sequence, with some divergence at the extreme ends of the colour range, where there are few stars to constrain the model fit.

The offset in the case of the Pleiades was caused by the long-standing discrepancy between Hipparcos and photometric methods. This is not present in the case of the Hyades, where Hipparcos-based distances generally agree with other methods.

Dispersion around the main sequence has been greatly reduced compared with computing the absolute magnitude from the data directly, by computing posterior distances for each star from the results of the fitting and the individual Hipparcos observations.

\begin{table}\centering
   \begin{tabular}{@{}rrrcrrcrrcrr@{}}\toprule  
        Parameter       & Estimated   & Error \\ \midrule
    Distance (pc)     &  46.35      &   0.35     \\  
    U (kms$^{-1}$)     &  -42.24   & 0.11\\
    V (kms$^{-1}$)     &  -19.27   & 0.12 \\
    W (kms$^{-1}$)     &  -1.55   &  0.11\\
    $\sigma_{UVW}$ (kms$^{-1}$)  & 1.10  &0.05\\
    \bottomrule
    \end{tabular}
    \caption{Colour-independent results obtained from the method applied to Hyades stars in the new Hipparcos reduction.}
    \label{table:hyades1}
\end{table}

\begin{table*}\centering
   \begin{tabular}{@{}rrrcrrcrrcrr@{}}\toprule  
    &  \multicolumn{2}{c}{ $(0.1<V-I<0.6)$ }  & \phantom{} & \multicolumn{2}{c}{ $(0.6<V-I<0.8)$ } & \phantom{} & \multicolumn{2}{c}{$(0.8<V-I<1.25)$} & \phantom{} & \multicolumn{2}{c}{$(1.25<V-I<1.8)$}\\  
           \cmidrule{2-3} \cmidrule{5-6} \cmidrule{8-9}  \cmidrule{11-12}
        Parameter       & Estimated    & Error & & Estimated    & Error & & Estimated    & Error & & Estimated    & Error\\ \midrule
 $\sigma_R$ (pc)   & 3.42  & 0.27   && 5.50  &  0.67   && 2.30    &  0.15  && 2.55   & 0.45  \\ 
 $\sigma_M$ (mag)         & 0.39  &  0.04  && 0.32  &  0.04  &&  1.63   & 0.19  && 0.34  & 0.12 \\
 A (start point)     & 0.77 &  0.24    && 3.98  &  0.05 && 5.25  &  0.07 && 7.25  & 0.15 \\
 B (end point)       & 3.98 & 0.05   &&  5.25  &  0.07  && 7.25  &  0.15 && 9.26  &  0.57 \\
   \bottomrule
    \end{tabular}
    \caption{Colour-dependent results obtained from the method applied to the new Hipparcos reduction for the Hyades open cluster. In the four bins there are 76, 55, 53, and 21 stars (low $(V-I)$ to high). }\label{table:hyades2}.  
\end{table*}

\section{Outlook for Gaia}
\label{sec:OutlookForGaia}
The method described in Sect. \ref{sec:math} will be particularly useful after the release of the Gaia astrometric catalogue. That the Gaia catalogue will include all of the information required for applying this method, including radial velocities for $G_{RVS}<$17, in one self-consistent catalogue makes Gaia ideal for studying open clusters to greater precision and at greater distances than was possible previously. Indeed, Gaia is expected to observe some one billion stars, including stars in numerous open clusters. This will allow application of this method to many more clusters, including those at much greater distances than was previously possible.

To test the performance of the ML method with Gaia data, a simulated open cluster was created, and simulated Gaia observational errors applied. To continue with the case study of the Pleiades, a simulated star catalogue for a Pleiades-like cluster was obtained from GaiaSimu. This (\cite{masana}, \cite{GUMS})  is a set of libraries containing the Gaia Universe Model and instrument models used by the Gaia Data Processing and Analysis Consortium (DPAC). It contains a database of 500 simulated open clusters, including the Pleiades, constructed from Padova isochrones and a Chabrier/Salpeter IMF. 

To apply simulated Gaia observational errors, the data was processed using the Gaia Object Generator (GOG). It \citep{GOG} is a simulator of the Gaia end of mission catalogue, and is one of the major products of DPAC's simulation efforts. It contains all of the currently available predicted error models for the Gaia satellite, and is capable of transforming an input catalogue of `true' stellar properties into simulated Gaia observations including predicted observational effects and the instrumental capabilities.

In the Gaia case, the selection function in Eq. \ref{eqn:HipLim} is modelled as a step function at $G=20$ mag. Because Gaia is expected to be complete up to this magnitude, the step function should be a good approximation to the real case. 

\subsection{Pleiades with Gaia}

The GaiaSimu and GOG simulated Pleiades contains some 1000 stars, placed at a distance of 130 pc and occupying the same region of the sky as the real Pleiades. With simulated parallax errors of between 10 and 100 $\mu$as, the vast majority of the star's distances are very accurately measured. 

With such a precise data set, both the estimated distance from MLE and the distance obtained directly from the mean of the parallax are both within 0.01 pc of the true value. This highlights that, for the nearest open clusters, it will be possible
with Gaia data to go further beyond the current goal of determining the distance and kinematic and structural parameters, to having highly detailed information on many aspects of open clusters. 

In such cases, the mean distance of a cluster determined through the ML method will not in itself be useful, although individual stars' posterior distance estimates from the method will be unbiased and therefore preferable to distances found by inverting individual parallaxes. In terms of determining a cluster's spatial distribution, the direct use of parallax information results in a bias in the results not present during the application of the ML method.

\subsection{Distant clusters with Gaia}

\begin{figure*}
\begin{center}
\includegraphics[width=\linewidth]{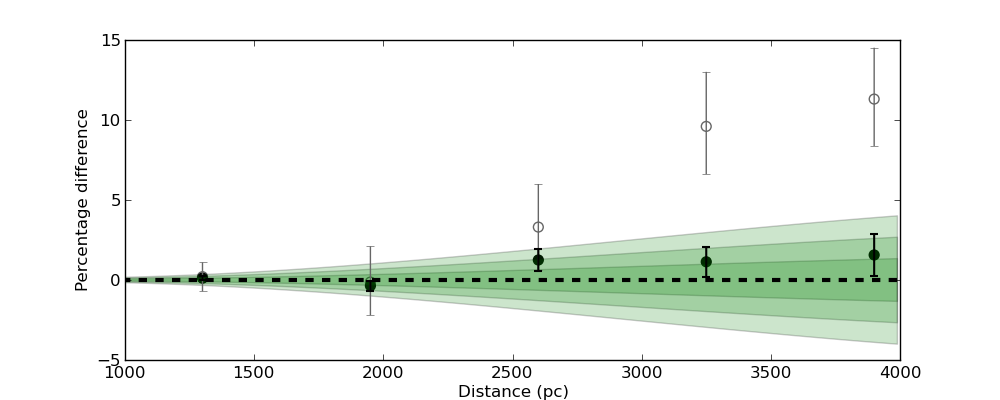} 
\caption{Results of the distance estimation to a simulated Pleiades-like cluster placed at a factor of 10 to 30 times the original distance. Filled circles show the percentage difference between the MLE-estimated distances and the true distance, open circles are the percentage difference between the inverse of the mean of the parallax and the true distance, and green shading highlights 1, 2, and 3$\sigma$ errors extrapolated over the entire range.}
\label{fig:GogResults}
\end{center}
\end{figure*}

To test the performance of the ML method with open clusters at greater distances, the GaiaSimu simulated Pleiades was modified, increasing the distance while conserving all other properties. The open cluster was moved to a range of distances between 10 and 30 times the originally assumed distance of 130 pc (i.e. up to a distance of 4 kpc). Then the Pleiades-like clusters were processed with GOG to simulate Gaia observational errors. 

Using the same simulated open cluster moved to different distances allows a direct comparison of the ML method performance at different distances.
Figure \ref{fig:GogResults} shows the results of the distance estimation using the ML method, showing the percentage difference between the `true' distance and the distance estimated from MLE, $\Delta(d_{real}-d_{MLE})/d_{real}$, and comparing this with the percentage difference between the `true' distance with the inverse of the mean parallax, $\Delta(d_{real}-d_{1/\bar{\varpi}})/d_{real}$. As the distance to the cluster increases, the stars become fainter and the observational errors larger. As can be clearly seen in Fig. \ref{fig:GogResults}, the mean of the parallax is susceptible to large random error, in addition to Lutz-Kelker effects and other statistical biases, and is unsuitable for accurate distance determination in magnitude-limited data sets and those with significant observational errors. 

\begin{figure}
\begin{center}
\includegraphics[width=\linewidth]{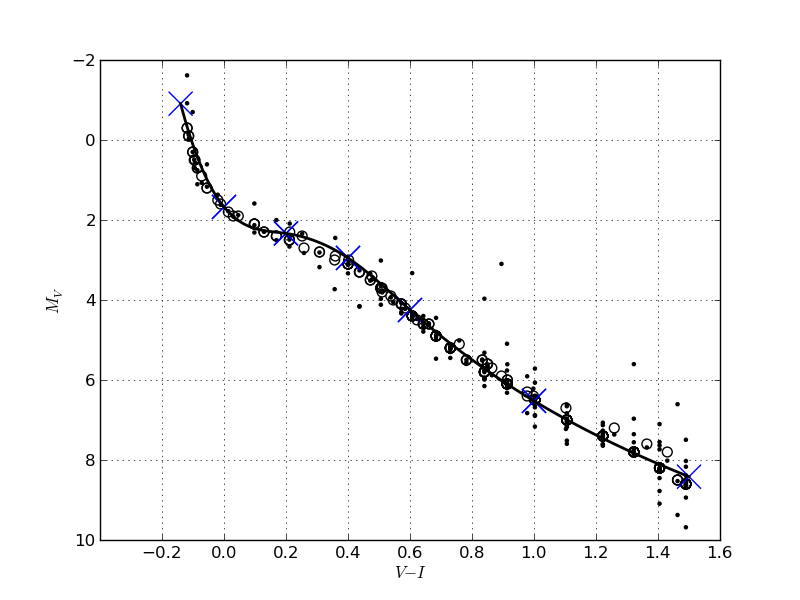} 
\includegraphics[width=\linewidth]{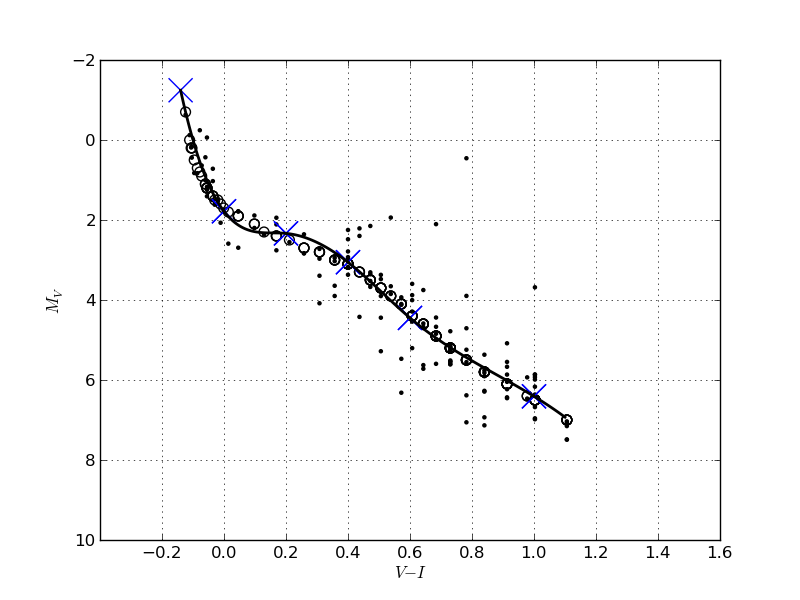} 
\includegraphics[width=\linewidth]{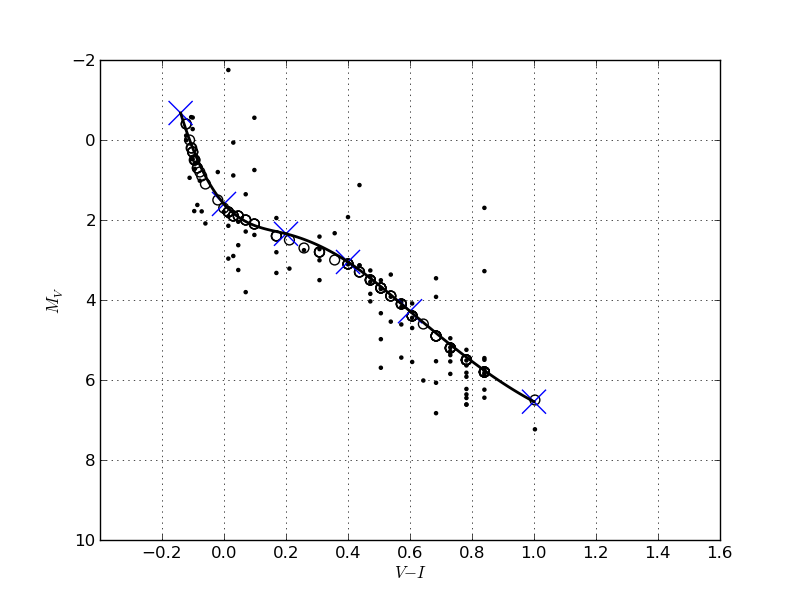} 
\caption{CMD for the simulated Pleiades like cluster at distances of 1300 pc (top), 2600 pc (middle) and 3900 pc (bottom). The three clusters have 216, 143 and 114 observed members respectively. Open circles show the `true' simulated absolute magnitudes without errors, and filled circles show the absolute magnitudes calculated directly from the simulated observations including simulated gaia errors. Stars with negative parallaxes are omitted from the figure but included in the ML estimation. Reddening in $V-I$ is assumed known. The blue crosses are the points fitted with the ML method and the solid line is the resulting isochrone-like sequence. }
\label{fig:GogCMD}
\end{center}
\end{figure}

As with the Hipparcos Pleiades and Hyades data, the CMD is plotted with the isochrone-like sequence obtained from the observational data and shown in Fig. \ref{fig:GogCMD}. These plots have been created using the simulated clusters at 1300, 2600, and 3900 pc, showing that it is possible to obtain a reliable observational isochrone from Gaia observations even when individual parallaxes are strongly affected by observational errors. 

With this simulated dataset, the ML method is confirmed as not suffering from significant statistical biases, and it is expected to perform well with real Gaia data.

\subsection{Membership selection}

When studying open clusters, especially those at greater distances, membership selection has always been important and problematic. When applying the ML method to distant open clusters in the Gaia catalogue, the density of stars on the sky could cause problems with misclassification and source confusion.

However, with an expected source density in the galactic plane of around 3$\times 10^5$ stars per square degree, Gaia's windowing system for object detection is small enough to give a low probability of source confusion even when observing distant open clusters.

In terms of membership selection, the ML method's estimation of an open cluster's parameters can be used directly to perform membership probability tests. If the ML method is primed using a sample of probable members, a $\chi^2$ test can be applied to calculate membership probability for each star in the sample using the parallax, proper motion, and radial velocity information simultaneously:

\begin{equation} 
K^2 = \textbf{A} C^{-1} \textbf{A}^{T}
\label{eqn:chisquared}
,\end{equation}
where $\textbf{A}$ is the vector $(\varpi_i-\varpi_{ML},\,\, \, \mu_{\alpha,i} - \mu_{\alpha,ML}, \,\,\, \mu_{\delta,i} - \mu_{\delta,ML},\,\, \, v_{ri} - v_{r,ML})$, and  $C^{-1}$ is the sum of the catalogue's covariance matrix and the variance on each parameter due to the clusters intrinsic dispersion in distance, proper motion, and radial velocity. Here, $\varpi_{ML}$, $\mu_{\alpha,ML}$, $\mu_{\delta,ML}$, and $v_{r,ML}$ are the mean parallax, mean cluster proper motion and mean radial velocity of the cluster, determined from the ML method.

Stars with $K^2 > 16.25$ are rejected, in correspondence with a $\chi^2$ test at three-sigma level with four degrees of freedom. This process should be performed using an iterative process, rejecting the worst outliers a few at a time and recalculating all fitting parameters, until no further outliers remain.

Here, the estimated cluster distance and space velocity, including intrinsic dispersion, are combined with the individual observations and their associated errors in order to distinguish between members and non-members in a single step.

Testing using a 1000-star sample of GOG simulated field stars added to the five simulated Pleiades samples used in Fig. \ref{fig:GogResults}, the $\chi^2$ test excluded field stars with a misclassification rate between 0.8 and 0.3\%, assuming a worst case of zero radial velocity information.

\section{Conclusions}
\label{sec:discussion}
An improved method for estimating the properties of open clusters has been presented, and tested using real data on two nearby and well studied open clusters. In addition to distance estimation, internal kinematics and spatial structure were probed, with mass segregation detected in the case of the Pleiades. These results confirm that the method performs as expected and highlight the potential future uses of such a method when high quality parallax information is available from the Gaia mission.

After revisiting the `Pleiades problem', we find that an explanation cannot be found in error correlation problems in Hipparcos. Through the use of simulations we find that Gaia will measure the distance to Pleiades stars with precision of a fraction of a percent, enabling a conclusion to this long running discrepancy.

The ML method can be extended further to give more detailed information, such as including a model for cluster ellipticity and orientation. It is possible to include and compare different spatial and kinematic distributions, allowing one to test predictions on spatial structure, mass segregation, and peculiar motions, and to test for other properties such as cluster rotation. In the case of the absolute magnitude distribution, it would be possible to give age and metallicity estimates by fitting and comparing sequences of different theoretical isochrones. 

Unresolved binaries, which complicate studies of open clusters, can be detected using the posterior distances calculated using the ML method and the resulting colour-magnitude diagram. It is possible to extend the method to use a distribution in absolute magnitude that is asymmetrical around the main sequence, in order to consider undetected unresolved binaries within the method.

As mentioned in Sect. \ref{sec:pleiades}, a lack of quality radial velocity data for Hipparcos Pleiades stars limits the application of the method in fitting the full three-dimensional kinematics of the open cluster. This is expected to change when Gaia comes to fruition, because all stars with $G_{RVS}<17$ will have radial velocity information from the on-board radial velocity spectrometer. Additionally, very high quality radial velocities for stars in more than 100 open clusters will become available through the Gaia ESO Survey \citep{GES}, expanding the scope of the method's application to clusters at much greater distances.

\begin{acknowledgements}

The authors would like to thank Carme Jordi for her useful comments, and Anthony Brown and Jos de Bruijne for their input on recreating the results of \cite{HipCorrelatedErrors}. We likewise thank CU2 for the use of GaiaSimu and the GOG simulator, and Erika Antiche for running GOG simulations.

This work was supported by the Marie Curie Initial Training
Networks grant number PITN-GA-2010-264895 ITN ``Gaia
Research for European Astronomy Training'', and MINECO (Spanish Ministry of Economy) - FEDER through grant AYA2009-14648-C02-01, AYA2010-12176-E, AYA2012-39551-C02-01, and CONSOLIDER CSD2007-00050. 

\end{acknowledgements}

%-------changed by A&A language editor ------- 
\bibliography{References} 
%---------------------------------------------

\begin{appendix}

\section{Cartesian to galactic coordinate transformation}
\label{Sec:CoordinateConversion}
The spatial distribution of the cluster is given in Cartesian coordinates by a Gaussian distribution in each axis:
\begin{equation} 
\varphi_{x} = e^{-0.5 \left( \frac{x-X_0}{\sigma_{S} }\right)^2} 
\end{equation}

\begin{equation} 
\varphi_{y} = e^{-0.5 \left( \frac{y-Y_0}{\sigma_{S} }\right)^2} 
\end{equation}

\begin{equation} 
\varphi_{z} = e^{-0.5 \left( \frac{z-Z_0}{\sigma_{S} }\right)^2} 
,\end{equation}

where $X_0$, $Y_0$, and $Z_0$ define the centre of the cluster in each axis, and $\sigma_{S}$ gives the variance of the distribution.

To transform this Cartesian PDF into polar coordinates as required for our observables $r$, $l$, and $b$, we start with the relationship between our two sets of variables and find the inverse:
\begin{equation}
r=\sqrt{x^2 + y^2 + z^2} \Longrightarrow   x=\sqrt{r^2-y^2-z^2}  
\end{equation}

\begin{equation} 
l = \textrm{tan}^{-1} \left( \frac{y}{x} \right) \Longrightarrow y=\textrm{tan}(l)x    
\end{equation}

\begin{equation}
b=\textrm{sin}^{-1} \left( \frac{z}{r} \right) \Longrightarrow \boxed{ z= r \textrm{sin}(b) }
.\end{equation}

Then we have two equations and two unknowns:

\begin{equation} 
x^2 = r^2 - y^2 - z^2 = r^2 - \textrm{tan}^2(l) x^2 - r^2 \textrm{sin}^2(b) 
\end{equation}

\begin{equation} 
x^2 (1+ \textrm{tan}^2(l)) = r^2- r^2 \textrm{sin}^2(b) 
\end{equation}

\begin{equation}
x^2 = \frac{r^2 - r^2 \textrm{sin}^2(b) }{1+ \textrm{tan}^2(l)} 
\end{equation}

\begin{equation}  
\boxed{x= r  \textrm{cos}(b) \textrm{cos}(l)} 
\end{equation}

\begin{equation} 
\boxed{y= r  \textrm{cos}(b) \textrm{sin}(l)} 
.\end{equation}

Then we require the Jacobian: \(r^2 \textrm{cos}(b) \).

By substituting the $x$, $y$, and $z$ found above into the original PDF and multiplying by the Jacobian of the transformation, we find the PDF in the new coordinate system:

\begin{equation} 
\begin{split}
\varphi_{rlb} = \\
&r^2 \textrm{cos}(b) \\
& e^{ - \frac{0.5}{\sigma_S^2} \left( (r \textrm{cos}(b)\textrm{cos}(l) - X_0 )^2 + (r\textrm{cos}(b)\textrm{sin}(l) -Y_0)^2 + (r\textrm{sin}(b)-Z_0)^2 \right) }   
\end{split}
.\end{equation}

By rotating our coordinate system $l,b\rightarrow l',b'$ to align the cluster centre with the X axis, we have new coordinates $l'$ and $b'$ for all the stars. In this rotated coordinate system, the cluster has a position $Y_0'=Z_0'=0$, and $X'$ is equivalent to the distance to the clusters centre. The above spatial probability distribution function can then be simplified as

\begin{equation}  
\varphi_{r'l'b'} = r^2 \textrm{cos}(b')  e^{ - \frac{0.5}{\sigma_S^2} \left(R^2 + r^2 -2rR \textrm{cos}(b') \textrm{cos}(l') \right)} 
\end{equation}
where $R$ is the distance to the cluster. 

It should be noted that the two coordinate systems are used simultaneously. The rotated coordinates $(l',b')$ are used in the integration over position to simplify the integrals as explained above. However, in the analytic solution to the integrals over $\mu_{\alpha^\ast} \mu_{\delta} v_r $, the unrotated coordinates $l$ and $b$ are used.

\section{Integration of the likelihood function}
\label{sec:IntegrationOfLikelihoodFunction}
To evaluate $\mathcal{D}(\mathbf{y}|\boldsymbol{\theta})$ in Eq. \ref{eqn:MLBasic} we must integrate over all $\mathbf{y}_0$, giving a multiple integral that can be split into three parts. First is the integral over variables with assumed zero error, second kinematics, and finally distance.
\subsection{Integration over $m_{0}$, $l_0'$ and $b_0'$}
As these variables have errors given by the delta function, $(m,l',b')=(m_{0},l_0',b_0')$ so we can use $(m,l',b')$. This avoids integrating over these three parameters.
\subsection{Integration over $\mu_{\alpha^\ast,0}$, $\mu_{\delta,0}$ and $v_{r0}$}
The triple integral over $\mu_{\alpha^\ast,0}$, $\mu_{\delta,0}$ and $v_{r0}$ is

\begin{equation}
\int_{\forall\mu_{\alpha^\ast,0} \mu_{\delta,0} v_{r0}  }    \varphi_{v}(U,V,W)  \mathcal{E}(z|z_0)  \, d\mu_{\alpha^\ast,0} \, d\mu_{\delta,0} \, dv_{r0}
\end{equation}
where 
\begin{equation}   
\mathcal{E}(z|z_0) = e^{-0.5 \left( \frac{\mu_{\alpha^\ast}-\mu_{\alpha^\ast,0}}{\epsilon_{\mu_{\alpha^\ast}}}\right) ^2 }   e^{-0.5 \left( \frac{\mu_{\delta}-\mu_{\delta,0}}{\epsilon_{\mu_{\delta}}}\right) ^2 } 
e^{-0.5 \left( \frac{v_r-v_{r0}}{\epsilon_{v_r}}\right) ^2 } 
.\end{equation}

In order to perform the integral, the function $\varphi_{v}(U,V,W)$ must be expressed in terms of $\mu_{\alpha^\ast,0}$, $\mu_{\delta,0}$, and $v_{r0}$. This is achieved through the following expressions:

 \begin{eqnarray}
U = a_1 \mu_{\alpha^\ast} r + b_1 \mu_{\delta} r + c_1 v_r  \nonumber \\
    a_1 = -k \textrm{cos}(b) \textrm{sin}(l)  \nonumber \\
    b_1 = -k \textrm{sin}(b) \textrm{cos}(l)  \nonumber \\
    c_1 = \textrm{cos}(b) \textrm{cos}(l)  \nonumber \\
& & \nonumber \\
 \end{eqnarray}
  \begin{eqnarray}
 V = a_2 \mu_{\alpha^\ast} r + b_2 \mu_{\delta} r + c_2 v_r \nonumber \\
   a_2 = k \textrm{cos}(b) \textrm{cos}(l)  \nonumber \\
  b_2 = -k \textrm{sin}(b) \textrm{sin}(l)  \nonumber \\
    c_2 = \textrm{cos}(b) \textrm{sin}(l)  \nonumber \\
& & \nonumber \\
 \end{eqnarray}
  \begin{eqnarray}
W = a_3 \mu_{\alpha^\ast} r + b_3 \mu_{\delta} r + c_3 v_r \nonumber \\
   a_3 = 0   \nonumber \\
   b_3 = k \textrm{cos}(b) \nonumber \\  
    c_3 = \textrm{sin}(b)   \nonumber \\
 \end{eqnarray}

where \( k = 4.74 \frac{Km \, year}{"s \, pc} \).

Therefore $\varphi_{v}(U,V,W)$ can be written in terms of $\mu_{\alpha^\ast,0}$, $\mu_{\delta,0}$ and $v_{r0}$ as
\begin{equation} \varphi_{v}(U,V,W)= e^{p(\mu_{\alpha^\ast,0}, \mu_{\delta,0}, v_{r0} |r)} .\end{equation} 
This can be integrated using the definite integral, 
\begin{equation}   
\int_{-\infty}^\infty e^{-(\alpha x^2 + \beta x + \gamma)} dx
       =
         \sqrt{\frac{\pi}{\alpha}} \;
         e^{ \left( \frac{\beta^2}{4 \alpha} - \gamma \right) } 
,\end{equation}
giving the solution
\begin{equation} 
\int_{\forall\mu_{\alpha^\ast,0} \mu_{\delta,0} v_{r0}  }    \varphi_{v}(U,V,W)  \mathcal{E}(\mathbf{y}|\mathbf{y}_0)  \, d\mu_{\alpha^\ast,0} \, d\mu_{\delta,0} \, dv_{r0} = K \; e^{\left( \frac{Y^2}{4 X} - Z \right) } 
\end{equation}
where $K$, $X$, $Y$, and $Z$ are defined in Eq. \ref{KXYZ}. 

\subsection{Integration over $R$}

The remaining integral has no analytical solution and will therefore be performed numerically:
\begin{equation}
\mathcal{D}(\mathbf{y}|\boldsymbol{\theta})= \int_{r} \varphi_{M_{0}}\varphi_{\varpi_0l_0'b_0'}
 K \; e^{\left( \frac{Y^2}{4 X} - Z \right) }  
 e^{-0.5 \left( \frac{\varpi-\varpi_0}{\epsilon_\varpi}\right) ^2 } \, dr_0
\end{equation}

 \begin{eqnarray}
          K & = & \sqrt{  - \frac{\pi^3}{J_2 E_1 X} } \nonumber \\
                      &   & \nonumber  \\    
          X & = & \frac{D_1^2}{4E_1} - F_1 \nonumber \\
          Y & = & \frac{B_1 D_1}{2E_1} - C_1 \nonumber \\
          Z & = & \frac{B_1^2}{4E_1} - A_1 \nonumber \\
            &   & \nonumber  \\
        A_1 & = & \frac{A_2^2}{4J_2} - D_2 \nonumber \\
        B_1 & = & \frac{A_2 B_2}{2J_2} - E_2 \nonumber \\
        C_1 & = & \frac{A_2 C_2}{2J_2} - F_2 \nonumber \\
        D_1 & = & \frac{B_2 C_2}{2J_2} - G_2 \nonumber \\
        E_1 & = & \frac{B_2^2}{4J_2} - H_2 \nonumber \\
        F_1 & = & \frac{C_2^2}{4J_2} - I_2 \nonumber \\
            &   & \nonumber  \\    
        A_2 & = & - \frac{\mu_{li}}{\epsilon_{\mu_{\alpha^\ast}}^2} 
                  - \left(\frac{a_1 U_0}{\sigma_U^2} 
                  + \frac{a_2 V_0}{\sigma_V^2} 
                  + \frac{a_3 W_0}{\sigma_W^2} 
                  \right) r  \nonumber \\    
        B_2 & = &   \left(
                           \frac{a_1 b_1}{\sigma_U^2} +
                           \frac{a_2 b_2}{\sigma_V^2} +
                           \frac{a_3 b_3}{\sigma_W^2}
                    \right) r^2 \nonumber \\      
        C_2 & = &   \left(
                           \frac{a_1 c_1}{\sigma_U^2} +
                           \frac{a_2 c_2}{\sigma_V^2} +
                           \frac{a_3 c_3}{\sigma_W^2}
                    \right) r \nonumber \\    
        D_2 & = & \frac{1}{2}
                  \left(
                         \frac{\mu_{li}^2}{\epsilon_{\mu_{\alpha^\ast}}^2} +
                         \frac{\mu_{bi}^2}{\epsilon_{\mu_{\delta}}^2} +
                         \frac{v_{ri}^2}{\epsilon_{v_r}^2}
                         \; + \;
                         \frac{U_0^2}{\sigma_U^2} +
                         \frac{V_0^2}{\sigma_V^2} +
                         \frac{W_0^2}{\sigma_W^2}
                  \right) \nonumber \\      
        E_2 & = & - \frac{\mu_{bi}}{\epsilon_{\mu_{\delta}}^2}
                  - \left(
                           \frac{b_1 U_0}{\sigma_U^2} +
                           \frac{b_2 V_0}{\sigma_V^2} +
                           \frac{b_3 W_0}{\sigma_W^2}
                    \right) r \nonumber \\ 
        F_2 & = & - \frac{v_{ri}}{\epsilon_{v_r}^2}
                  - \left(
                           \frac{c_1 U_0}{\sigma_U^2} +
                           \frac{c_2 V_0}{\sigma_V^2} +
                           \frac{c_3 W_0}{\sigma_W^2}
                    \right) \nonumber \\ 
        G_2 & = &   \left(
                           \frac{b_1 c_1}{\sigma_U^2} +
                           \frac{b_2 c_2}{\sigma_V^2} +
                           \frac{b_3 c_3}{\sigma_W^2}
                    \right) r \nonumber \\
        H_2 & = & \frac{1}{2}
                  \frac{1}{\epsilon_{\mu_{\delta}}^2}
                  \; + \;
                  \frac{1}{2}
                  \left(
                         \frac{b_1^2}{\sigma_U^2} +
                         \frac{b_2^2}{\sigma_V^2} +
                         \frac{b_3^2}{\sigma_W^2}
                  \right) r^2\nonumber \\
        I_2 & = & \frac{1}{2}
                  \frac{1}{\epsilon_{v_r}^2}
                  \; + \;
                  \frac{1}{2}
                  \left(
                         \frac{c_1^2}{\sigma_U^2} +
                         \frac{c_2^2}{\sigma_V^2} +
                         \frac{c_3^2}{\sigma_W^2}
                  \right) \nonumber \\
 J_2 & = & \frac{1}{2}
                  \frac{1}{\epsilon_{\mu_{\alpha^\ast}}^2}
                  \; + \;
                  \frac{1}{2}
                  \left(
                         \frac{a_1^2}{\sigma_U^2} +
                         \frac{a_2^2}{\sigma_V^2} +
                         \frac{a_3^2}{\sigma_W^2}
                  \right) r^2 \nonumber \\
\label{KXYZ}
\end{eqnarray}

\section{Normalisation coefficient}

Until now we have been using the un-normalised joint probability distribution. Normalisation is achieved by dividing by a normalisation constant, $ \mathcal{C}$. The normalisation constant is found by integrating the un-normalised joint probability distribution $\mathcal{D}(\mathbf{y}|\boldsymbol{\theta})$ over all $\mathbf{y}$:

\begin{equation} 
\mathcal{C} = \int_{\forall \mathbf{y}_0} \int_{\forall \mathbf{y}}\varphi_{M_{0}}\varphi_{\varpi_0l_0'b_0'}  \varphi_{v0}(U,V,W) \mathcal{S}(\mathbf{y}) \mathcal{E}(\mathbf{y}|\mathbf{y}_0) \, d\mathbf{y} \, d\mathbf{y}_0
.\end{equation}

This integral can be performed in two parts, where $I$ is defined such that\begin{equation} 
\mathcal{C} = \int_{\forall \mathbf{y}_0}\varphi_{M_{0}} \varphi_{\varpi_0l_0'b_0'}  \varphi_{v0}(U,V,W)  \underbrace{\int_{\forall \mathbf{y}}  \mathcal{S}(\mathbf{y}) \mathcal{E}(\mathbf{y}|\mathbf{y}_0) \, d\mathbf{y}}_\textrm{I} \, d\mathbf{y}_0
.\end{equation}

Substituting in the selection function $\mathcal{S}(\mathbf{y})$ and the PDF of the observational errors $\mathcal{E}(\mathbf{y}|\mathbf{y}_0)$ gives the following seven-dimensional integral:

\begin{equation}
 I= \int_{\forall \mathbf{y}} \boldsymbol{\theta}(m-m_{lim}) \mathcal{E}(\mathbf{y}|\mathbf{y}_0) \, d\mathbf{y}
.\end{equation}

This integral can be split into two parts. The integral over the delta function in $\mathcal{E}(\mathbf{y}|\mathbf{y}_0)$ that, by definition, gives one; and the integral over each Gaussian error,

\begin{equation}
\begin{split}
I = \\
&\boldsymbol{\theta}(m-m_{lim}) \\
&\int_{\forall \varpi \mu_{\alpha^\ast} \mu_{\delta} v_r}  e^{-0.5 \left( \frac{\varpi-\varpi_0}{\epsilon_\varpi}\right) ^2 }  e^{-0.5 \left( \frac{\mu_{\alpha^\ast}-\mu_{\alpha^\ast,0}}{\epsilon_{\mu_{\alpha^\ast}}}\right) ^2 } 
e^{-0.5 \left( \frac{\mu_{\delta}-\mu_{\delta,0}}{\epsilon_{\mu_{\delta}}}\right) ^2 } 
\\
&e^{-0.5 \left( \frac{v_r-v_{r0}}{\epsilon_{v_r}}\right) ^2 }    \,d\varpi \, d\mu_{\alpha^\ast} \, d\mu_{\delta} \,d v_r \int_{-\infty}^{\infty} \delta(m,l',b') dm \, dl' \, db'  
\end{split}
\end{equation}

\begin{equation} 
I= \boldsymbol{\theta}(m-m_{lim})  (2\pi) ^2  \epsilon_{\varpi} \epsilon_{\mu_{\alpha^\ast}} \epsilon_{\mu_{\delta}} \epsilon_{v_r}     
.\end{equation}
Here, $\boldsymbol{\theta}(m-m_{lim})$ acts to provide an upper limit to the integral over all $M_G$. Substituting $I$ back into $C$ we have

\begin{equation}
\begin{split}
 \mathcal{C} = \\
&    (2\pi) ^2  \epsilon_{\varpi} \epsilon_{\mu_{\alpha^\ast}} \epsilon_{\mu_{\delta}} \epsilon_{v_r}  \\
&\int_{-\infty}^{m_{\rm lim}} \int_0^{\infty} \int_{-\pi/2}^{\pi/2} \int_0^{2\pi}  \int_{-\infty}^{\infty} \int_{-\infty}^{\infty} \int_{-\infty}^{\infty} \varphi_{m0} \varphi_{r_0l_0'b_0'}\\
& \varphi_{v}(U,V,W)  \, dU \, dV \,dW  \, dl_0' \, db_0'  \,dr_0 \, dm_0  
\end{split}
.\end{equation}
As with in the previous section, the integral can be split up into a number of parts. 

\subsection{Integration over $M_G$}
Evaluating first the integral over apparent magnitude gives
\begin{equation} 
\int_{-\infty}^{m_{\rm lim}} \varphi_{m0} \, dm_0  =   \frac{\sqrt{2 \pi}}{2} \sigma_M   \textrm{erfc} \left( \frac{A-m_{\rm lim}}{\sqrt2 \sigma_M} \right) 
\end{equation}
where erfc is the complementary error function, and
\begin{equation}
A= 5\textrm{log}(r_0) -5 +M_{\rm mean} 
.\end{equation}

\subsection{Integration over $(U,V,W)$}

Integrating over (U,V,W) gives

\begin{equation} 
\int_{-\infty}^{\infty} \int_{-\infty}^{\infty} \int_{-\infty}^{\infty}  \varphi_{v}(U,V,W)  \, dU \, dV \, dW = (2\pi)^{3/2} \sigma_U \sigma_V \sigma_W 
.\end{equation}

\subsection{Integration over $l_0'$, $b_0'$, and $r_0$}

The remaining triple integral has no analytical solution and will be performed numerically:

\begin{equation}
\begin{split}
 \mathcal{C} =  \\
&B \int_0^{\infty} \int_{-\pi/2}^{\pi/2} \int_0^{2\pi}\textrm{erfc} \left( \frac{A-m_{\rm lim}}{\sqrt2 \sigma_M} \right)  \varphi_{r_0l_0'b_0'} \, dl_0' \, db_0'  \,dr_0   
\end{split}
\end{equation}

with: \( B =  \frac{(2\pi)^{4}}{2} \sigma_U \sigma_V \sigma_W \sigma_M  \epsilon_{\varpi} \epsilon_{\mu_{\alpha^\ast}} \epsilon_{\mu_{\delta}} \epsilon_{v_r} \)

\end{appendix}
\end{document}